\documentclass[epsfig,floats,prl,twocolumn,preprintnumbers,floatfix,superscriptaddress]{revtex4-1}
\usepackage{graphicx}
\usepackage[latin2]{inputenc}
\usepackage{amsmath}
\usepackage{amssymb}
\usepackage{bm}
\usepackage{color}

\newcommand{\ainit}{\alpha_{\circ}}
\newcommand{\tm}{\theta_\text{max}}
\newcommand{\smax}{\sigma_\text{m}}
\newcommand{\gmax}{\gamma_{_\text{m}}}
\newcommand{\sref}{\sigma_\circ}

\begin{document}
\title{Yielding and Strain Stiffening in Entangled Assemblies of Frictional Granular Chains}
\author{M.\ Reza Shaebani}
\affiliation{Department of Theoretical Physics and Center for Biophysics, 
Saarland University, 66123 Saarbr\"ucken, Germany}
\author{Mehdi Habibi}
\affiliation{Laboratory of Physics and Physical Chemistry of Foods, 
Wageningen University, 6708WG Wageningen, The Netherlands}

\begin{abstract}
Packings of macroscopic granular chains capture some of the essential 
aspects of molecular polymer systems and have been suggested as a 
paradigm to understand the physics on a molecular scale. However, here 
we demonstrate that the interparticle friction $\mu$ in granular chain 
packings, which has no counterpart in polymer systems, leads to a 
nontrivial yielding and rheological response. Based on discrete 
element simulations we study the nonlinear rheology of random packings 
of granular chains under large amplitude oscillatory shear. We find 
that the maximum stress and the penetration depth of the shear 
deformation into the material bulk are nonmonotonic functions of 
friction with extrema at intermediate values of $\mu$. We also show 
that the regularly repeated gaps between the adjacent grains, which 
are special to commercial granular chains, broaden the shear zone and 
enhance the entanglements in the system by promoting the interlocking 
events between chains. These topological constraints can significantly 
increase the degree of strain stiffening. Our findings highlight the 
differences between the physics of granular chain packings and molecular 
polymer systems.
\end{abstract}

\maketitle

Polymer assemblies are key components of a variety of natural and synthetic 
materials \cite{Rubenstein03}. Whereas many properties of these materials 
can be understood from the microscopic states and configurations of their 
constituent polymeric elements, it is technically very difficult to probe 
such microstructures directly. As a promising alternative, macroscale 
granular chains have been offered due to their analogies to polymer systems 
(e.g.\ chain stiffness- or length-dependence of the mechanical response, 
similar roles of packing fraction and temperature in jamming transition, 
etc.)\,\cite{Reichhardt09,Zou09,Safford09,Taheri18,Brown12,Dumont18}; but 
even the physics of macroscale disordered assemblies of long semiflexible 
objects has remained largely unexplored. The presence of topological 
constraints in these systems \cite{Brown12,Dumont18,Shaebani22,Sarate22,
Gomez20,BenNaim01,Lopatina11,Soh19} leads to intriguing phenomena, such 
as strain stiffening under shear \cite{Brown12,Dumont18,Shaebani22}, and 
makes them different from, e.g., packings of rods where particle elongation 
and volume exclusion govern the behavior. Understanding the yielding, flow, 
and stiffening of entangled athermal systems of semiflexible chains is 
crucial to uncover the underlying structure of natural filamentous 
assemblies, such as bird nests \cite{Hansell05}, and for design of 
disordered meta materials \cite{Aktas21,Verhille17,Mirzaali17,Weiner20} 
and new smart textiles \cite{Weiner20,Yun13,Sunami22,Hu12,Poincloux18}.

How different entanglement mechanisms--- such as interlocking events between 
chains \cite{Dumont18} or formation of semiloops \cite{Brown12}--- contribute 
to the overall mechanical response has remained unclear, which is crucial 
to develop a quantitative theory of yielding and stiffening of entangled 
chain packings. More importantly, the presence of frictional interactions 
increases the complexity of the problem on the macroscale compared to 
molecular polymer systems. While the impact of friction on the structure 
and dynamics of granular systems has been extensively studied \cite{Goldenberg05,
Shaebani07,Shaebani08,Unger05,Singh20,Ostojic06,Fall14,Farain22,Rahbari21}, 
a detailed understanding of how the interplay between topological constraints and 
interparticle friction governs the mechanical response of chain packings 
is currently lacking. Furthermore, much less is known about the nonlinear 
rheology of these systems when yielding under shear \cite{Regev13}. 
Rheological response of dissipative systems to an oscillatory shear 
has been of particular interest \cite{Parley22,Tapadia06,Baggioli20,
Kamani21}. How far the shear-induced deformation penetrates into the 
bulk of granular-chain packings and what role the key factors, 
particularly friction and entanglements, play are intriguing problems 
\cite{Shaebani22}.

\begin{figure}[b]
\centering
\includegraphics[width=0.46\textwidth]{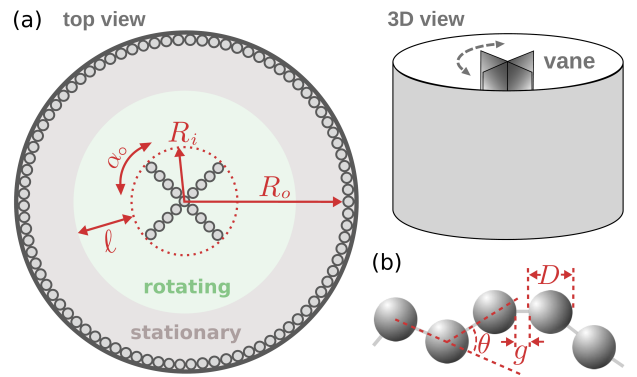}
\caption{(a) Schematic of the simulation setup. The four-blade vane of 
radius $R_i$ rotates with an amplitude $\ainit$ inside a cylindrical 
container of radius $R_o$ filled with granular chains. Background colors 
illustrate the rotating inner part and the stationary outer region, with 
the border defining the penetration depth $\ell$. (b) Illustration of 
the granular chains composed of spherical grains of diameter $D$. A 
constant gap $g$ is imposed (as a bond) between the neighboring grains 
along the chain. The bond angle $\theta$ can vary within the $0\,{\leq}
\,\theta\,{\leq}\,\tm$ range.}
\label{Fig1}
\end{figure}

Here, we numerically study the response of frictional granular-chain 
packings to large amplitude oscillatory shear (LAOS); see Fig.\,\ref{Fig1}. 
By measuring the steady-state maximum stress and the penetration depth of 
the shear deformation into the material bulk, we find that the mechanical 
and rheological responses of chain packings show nonmonotonic dependencies 
on the coefficient of friction $\mu$. The competition between a higher 
connectivity of the contact network at low $\mu$ and a larger local 
freedom for the tangential frictional forces at high $\mu$ leads to the 
highest stability at intermediate values of $\mu$. We also disentangle 
the contribution of interlocking and semiloops to the rheological 
response and show that the presence of regular gaps $g$ along commercially 
available granular chains can enhance the degree of strain stiffening.

We consider a packing of granular chains inside a cylindrical container 
where a rotating four-blade vane applies a sinusoidal deformation $\alpha\,
{=}\,\ainit \sin(2\pi f t)$. The rotation frequency is set to $f{=}\,0.1\,
\text{Hz}$ and the rotation amplitude satisfies $\ainit{\geq}10^{\circ}$ 
(corresponding to shear strains $\gamma{\gtrsim}0.17$) in all simulations to 
remain in the LAOS regime \cite{Shaebani22}. We use the contact dynamics 
(CD) method for rigid particles \cite{Jean99} to perform large-scale parallel 
simulations \cite{Shojaaee12}. The unit of the length is set to the diameter 
$D$ of the spherical grains which are used to construct the chains and rigid 
blades and to roughen the container surfaces to avoid slip. The radii 
of the blades and container are $R_i\,{=}\,4.5 D$ and $R_o\,{=}\,25 D$, 
respectively. To construct the chains we impose a gap $g$ between the surfaces 
of adjacent grains, as if there is a rigid bond between them; thus, the 
distance between the successive grain centers along the chain is $g{+}D$ as 
shown in Fig.\,\ref{Fig1}. The concept of introducing a fixed gap size is 
suited very well to the CD method where the interparticle forces are handled 
as constraint forces \cite{Jean99,Shaebani09}. We vary $g$ in our simulations 
but it is restricted to $g{\leq}\,0.4\,D$ to keep the adjacent grains close 
enough to each other such that the imaginary bonds never touch. The angle 
$\theta$ between the successive imaginary bonds is flexible to vary within 
the $0\,{\leq}\,\theta\,{\leq}\,\tm{=}\,40^{\circ}$ range. One can assign 
a local persistence length $\ell_p$ to the chain, deduced from the local 
directional persistence $p\,{=}\,\text{cos}(\theta)$ via $p\,{=}\,
\text{e}^{-\ell{/}\ell_p}$ \cite{Doi86,Shaebani20,Taheri18}.

\begin{figure}[t]
\centering
\includegraphics[width=0.46\textwidth]{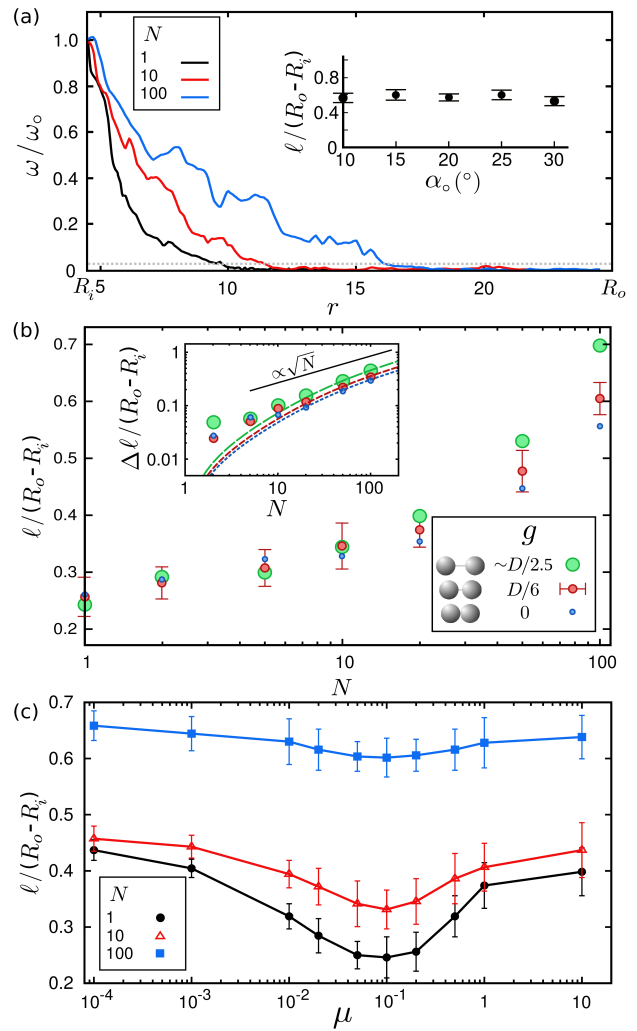}
\caption{(a) Mean angular velocity $\omega$, scaled by the mean angular 
velocity of the vane $\omega_{\circ}{=}\,2 f \ainit$, vs the radial distance 
$r$ from the cylinder axis (in units of $D$) for $\ainit{=}\,20^{\circ}$, 
$\mu\,{=}0.2$, and different chain lengths $N$. The dotted line denotes the 
threshold angular velocity $\omega_\text{c}{=}\,0.03\,\frac{\text{rad}}{\text{s}}$. 
Inset: Penetration depth $\ell$ vs the rotation amplitude $\ainit$ for $N
{=}100$. (b) $\ell$ vs $N$ for different values of the bond length $g$. The 
standard deviation is shown for $g{=}\frac{D}{6}$ data as an example. Inset: 
Excess penetration depth $\Delta\ell$ vs $N$. The dashed lines represent 
$\Delta\ell$ obtained via Eq.\,(\ref{Eq1}). (c) $\ell$ vs $\mu$ for $g{=}
\frac{D}{6}$ and different $N$.}
\label{Fig2}
\end{figure}

First, the shear cell is loaded with chains of equal length $N$ up to the 
height $h\,{\approx}\,25 D$ and the packing is relaxed into equilibrium 
under gravity. We construct a new packing for each value of the interparticle 
friction $\mu$. The oscillatory shear deformation in the LAOS regime is 
then performed and the stress and strain are measured after each 
$0.1^{\circ}$ change in the deflection angle. We previously proved that 
the numerical predictions of our simulation method are in quantitative 
agreement with the experimental results under similar conditions 
\cite{Shaebani22}. 

\begin{figure*}[t]
\centering
\includegraphics[width=0.98\textwidth]{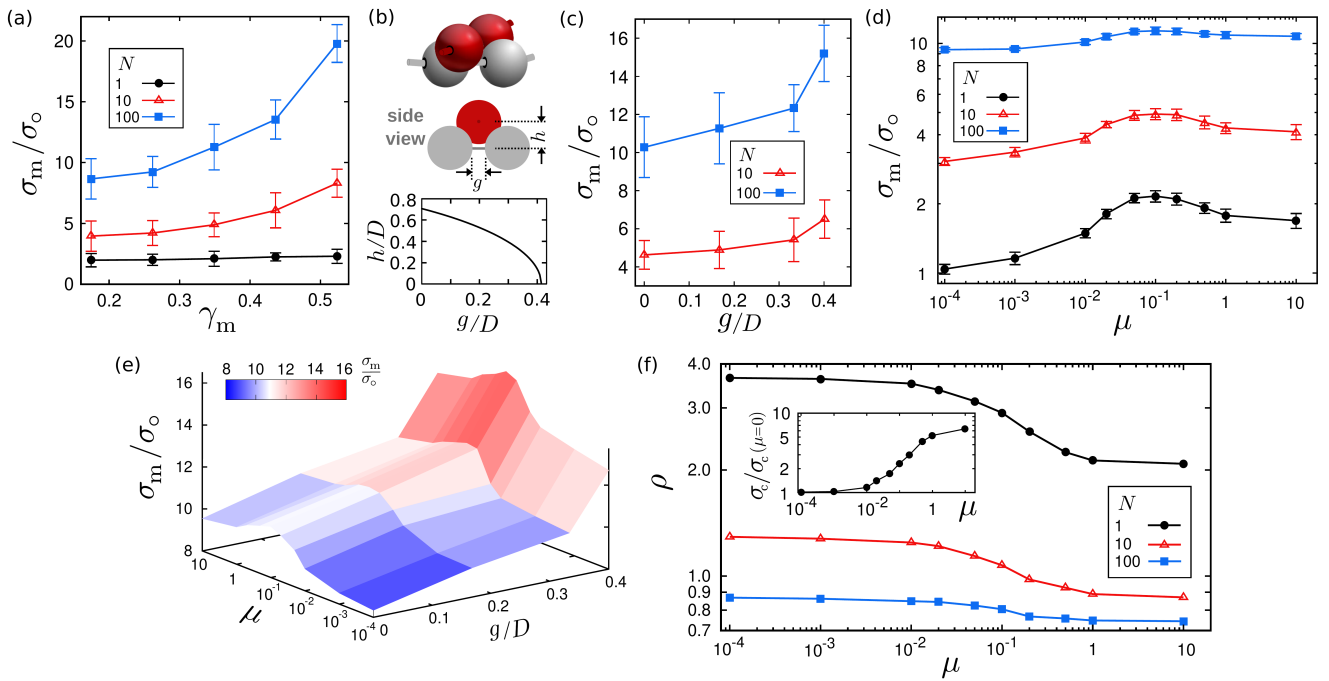}
\caption{(a) Maximum normal stress $\smax$ exerted on the vane, scaled by 
the similar stress $\sref$ in a packing with $\mu\,{=}0$ and $N{=}1$, vs 
the maximum shear strain $\gmax ({\propto}\,\ainit)$ for $\mu\,{=}0.2$ and 
different chain lengths. (b) Illustration of an interlocking event between 
the grains of different chains (top) and the minimum possible distance $h$ 
between the bonds vs $g$ (bottom). (c) $\smax$ vs $g$ for $\mu\,{=}0.2$, 
$\ainit{=}\,20^{\circ}$, and different $N$. (d) $\smax$ vs $\mu$ for $g{=}
\frac{D}{6}$ and different $N$. (e) $\smax$ in the ($g, \mu$) space for 
$N{=}100$. (f) Contact density $\rho$ (in units of $D^{-3}$) vs $\mu$ for 
different $N$. Inset: Critical normal stress $\sigma_\text{c}$ exerted on 
the vane at the onset of yielding, scaled by $\sigma_\text{c}$ for $\mu\,
{=}0$, vs $\mu$. The friction is changed before shearing a static packing 
of chains with $N{=}100$ constructed at $\mu\,{=}\,0$.}
\label{Fig3}
\end{figure*}

{\it Penetration depth of the deformations.---} While the material at the 
central part of the cell rotates with the blades, the movements beyond the 
blade edges decay with the distance from the moving boundary. Typical angular 
velocity profiles shown in Fig.\,\ref{Fig2}(a) indicate that increasing the 
chain length $N$ leads to larger fluctuations of the velocity profile, since 
the yielding domain depends stronger on the initial packing configuration. 
Moreover, increasing $N$ enhances the size of the shear zone. Wide shear 
zones away from the boundaries were previously reported in granular materials 
\cite{Fenistein03,Moosavi13,Shaebani21}; in contrast, here a wide shear zone 
forms near the moving boundary. We characterize the size of the shear zone 
by the penetration depth $\ell$, defined as the radial distance from the blade 
edges at which the angular velocity falls below a threshold value $\omega_\text{c}$ 
(i.e.\ $0\,{\leq}\,\ell\,{\leq}\,R_o{-}R_i$); see Figs.\,\ref{Fig1} and 
\ref{Fig2}(a). Here, the results are presented for $\omega_\text{c}{=}\,
0.03\,\frac{\text{rad}}{\text{s}}$ but our reported trends and conclusions 
are fairly insensitive to this choice. By varying $\ainit$ from $10^{\circ}$ 
to $30^{\circ}$ and repeating the simulations at each $\ainit$ for different 
initial configurations, we make sure that $\ell$ is independent of $\ainit$ 
in the LAOS regime; see, e.g., the inset of Fig.\,\ref{Fig2}(a). 

The dependence of $\ell$ on $N$ is shown in Fig.\,\ref{Fig2}(b). The behavior 
at the small $N$ regime is mainly dominated by the rheology of individual 
grain systems (i.e.\ $N{=}1$) and $\ell$ grows slowly with $N$; however, the 
growth rate increases for long chains. It can be also seen that $\ell$ varies 
independently of $g$ for small $N$ but systematically increases with $g$ at 
the large $N$ regime. It is known that stress reduction at shear-direction 
reversals in oscillatory shear of granular assemblies slightly broadens the 
shear zone up to a few grain diameters \cite{Toiya04}. However, formation 
of wider shear zones in our chain packings is due to the correlated dynamics 
of rotating chains. For sufficiently long chains, we hypothesize that the 
penetration depth of the shear deformation roughly scales with the length-scale 
that the chains in contact with the blade edges extend into the bulk of the 
system. By calculating the mean end-to-end distance $\xi$ of a flexible chain 
of length $N$ with segment length $D{+}g$ and a mean persistence $p$ in the 
{\it Supplemental Material} (see also persistent random walk models 
\cite{Tierno16,Nossal74}) and assuming that $\ell\,{\propto}\,\xi$, we obtain
\begin{equation}
\ell \propto \Big[\frac{1{+}p}{1{-}p}N + \frac{1}{(1{-}p)^2} (2p^N{+}p^2{-}
2p{-}1)\Big]^{0.5}\!\!(D{+}g)
\label{Eq1}
\end{equation}
for $N{>}1$, which reduces to $\ell\,{\propto}\,\sqrt{\frac{1{+}p}{1{-}p}N}(
D{+}g)$ in the large $N$ limit \cite{Doi86}. After subtracting the $\ell$ of 
individual grains, we obtain the excess penetration depth $\Delta\ell\,{=}\,
\ell(N)\,{-}\,\ell(N{=}1)$, as the pure contribution from the chain structure. 
The inset of Fig.\,\ref{Fig2}(b) shows that $\Delta\ell$ extracted from 
Eq.\,(\ref{Eq1}) satisfactorily captures both $N$ and $g$ dependence of 
$\ell$ in the large $N$ limit.

Next, we compare $\ell$ in packings constructed with different coefficients 
of friction $\mu$. Our striking finding is that $\ell$ develops a minimum at 
$\mu\,{\approx}\,0.1$ for all chain lengths [see Fig.\,\ref{Fig2}(c)]. Thus, 
the packings constructed with intermediate values of $\mu$ resist stronger 
against yielding, leading to smaller shear zones. The effect, however, 
becomes less pronounced in the large $N$ limit.

{\it Mechanical response and strain stiffening.---} By measuring the maximum 
normal stress $\smax$ exerted on the vane, we study the role of the key factors 
on the stress development in granular chain systems. The yield stress 
of individual grain packings varies negligibly with the rotation amplitude 
in the LAOS regime, as shown in Fig.\,\ref{Fig3}(a). In contrast, in 
chain packings with $N{\geq}9$ (i.e.\ the minimum length required for 
the formation of full rings), increasing the shear strain strengthens 
the entanglements through semiloop formation, leading to a pronounced 
strain stiffening \cite{Brown12,Shaebani22}. However, the developed 
stress is partially due to the interlockings between the grains of 
different chains. To evaluate the relative contributions of these two 
types of topological constraints (i.e.\ semiloops vs interlocking events), 
we vary the minimum possible distance $h$ between the bonds. This is 
achieved by varying the gap size $g$, which is related to $h$ via $h{=}
\sqrt{(D^2{-}2gD{-}g^2)/2}$; see Fig.\,\ref{Fig3}(b). The strength of 
each interlocking entanglement event grows with increasing $g$ (decreasing 
$h$), as it becomes more difficult for the grains to move over each 
other. Figure\,\ref{Fig3}(c) shows that $\smax$ can grow even up to 
$50\%$ upon varying $g$ from $0$ to ${\sim}\,0.4D$ (We remain in the 
range $g{\leq}g_\text{max}{=}(\sqrt{2}{-}1)D$; otherwise the bonds 
can touch, which further complicates the behavior). Notably, longer 
chains develop a larger degree of strain stiffening upon increasing 
the bond length. Indeed in the large $N$ limit the majority of the 
interlocking events occur between different chains: The volume 
covered by a chain and the volume of its constituent grains scale 
with $\xi^3{\sim}N^{1.5}$ and $N$, respectively; thus, the occupation 
fraction for each chain decays as ${\sim}1{/}\sqrt{N}$ and the chain 
rarely intersects itself in the limit $N{\rightarrow}\infty$. 

Similar to the penetration depth, the maximum stress $\smax$ exhibits 
a nonmonotonic behavior with the coefficient of friction. Figure\,\ref{Fig3}(d) 
shows that $\smax$ reaches a maximum at $\mu\,{\approx}\,0.1$, which 
indicates again that the assemblies constructed with intermediate 
values of friction are more stable structures and resist stronger 
against shear deformations. Figure\,\ref{Fig3}(e) summarizes the 
stress response of a granular chain packing in the ($g, \mu$) space.

\begin{figure}[b]
\centering
\includegraphics[width=0.48\textwidth]{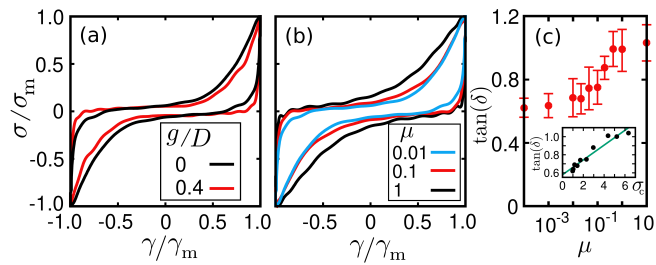}
\caption{Elastic Lissajous plots for $N{=}100$, $\gmax{\simeq}\,0.35$ 
and (a) $\mu\,{=}\,0.01$ and different bond lengths, and (b) $g{=}0$ 
and different values of $\mu$. (c) Loss tangent vs $\mu$ in packings 
with $g{=}0$. Inset: $\tan(\delta)$ vs yield stress $\sigma_\text{c}$. 
The line represents a linear fit.}
\label{Fig4}
\end{figure}

To understand the intriguing behavior as a function of $\mu$, the 
crucial point to note is that each value of $\mu$ is accompanied 
by a different granular-chain packing structure because the packings 
are newly generated for each value of $\mu$ (These structures differ 
from the frictional fiber assemblies generated by other protocols 
\cite{Ward15,Negi21}). It is known that frictional random packings 
of individual grains are hyperstatic structures in which the mean 
coordination number $z$ undergoes a transition with $\mu$ \cite{Unger05,
Shaebani09b,Silbert10,Wang10}. Since the stress tensor components 
in a random granular packing are proportional to the density of 
the interparticle contacts $\rho{=}z \phi{/}2 V_g$ (with $\phi$ 
and $V_g$ being the packing fraction and grain volume, respectively) 
\cite{Shaebani12,Shaebani12b}, here we measure $\rho$ as a function 
of $\mu$ in chain packings. As shown in Fig.\,\ref{Fig3}(f), $\rho$ 
exhibits plateaus at high and low friction limits and shows a smooth and 
monotonic transition in between. We hypothesize that the nonmonotonic 
mechanical response results from two competing effects: while increasing 
$\mu$ lowers the connectivity of the contact network, it strengthens 
the structure by providing larger local freedom for the tangential 
forces. The $\rho\,$-$\mu$ transition is shallower in assemblies 
with longer chains, which explains why the non-monotonicity of 
the mechanical response is less pronounced in the large $N$ limit. 
In support of our hypothesis, we measure the response of the same 
structure for different values of $\mu$. In this way, the role of 
connectivity is switched off and $\mu$ should solely strengthen 
the response. We take a static frictionless packing of chains with 
$N{=}100$, change the value of friction, and exert an increasing 
normal stress on the vane to bring the packing to the onset of yielding. 
By repeating this procedure for different values of $\mu$, we find 
that the critical yield stress $\sigma_\text{c}$ exerted on the 
vane grows monotonically with $\mu$ [inset of Fig.\,\ref{Fig3}(f)].

{\it Nonlinear rheology.---}
After applying many cycles of oscillatory shear to reach the steady state, 
we analyze the stress-strain relation in one shear cycle to clarify how 
the evolution of the contributions of shear strain $\gamma$ and shear 
rate $\dot{\gamma}$ depend on $g$ and $\mu$. The elastic Lissajous curves 
presented in Fig.\,\ref{Fig4} show that the rheological response is highly 
nonlinear, i.e., higher harmonics are present in the signal. The degree of 
intercycle stiffening slightly increases with $g$ [see Fig.\,\ref{Fig4}(a)]; 
thus, elastic stresses grow with the bond length. In contrast, viscous 
stresses develop with increasing $\mu$ leading to wider Lissajous curves 
[Fig.\,\ref{Fig4}(b)]. This can be quantified by the dimensionless loss 
tangent, $\tan(\delta)$, which represents the ratio of dissipated to stored 
energy in one cycle. Figure\,\ref{Fig4}(c) reveals that the rate of energy 
dissipation first increases with $\mu$ but slows down in the large $\mu$ 
limit, presumably due to weak network connectivity at large coefficients 
of friction. Interestingly, $\tan(\delta)$ is highly correlated to the 
yield stress [Fig.\,\ref{Fig4}(c)-inset], suggesting that the yield point 
can be predicted from, e.g., the complex shear modulus.  

In summary, our results demonstrate that the physics of packings of entangled 
granular chains differs from that of molecular polymer systems. The intriguing 
role of friction in complex assemblies of long semiflexible objects can have 
possible implications for design of new smart textiles and disordered meta 
materials \cite{Aktas21,Verhille17,Mirzaali17,Weiner20,Weiner20,Yun13,Sunami22,
Hu12,Poincloux18}. Understanding the structure and dynamics of externally-driven 
interacting filamentous assemblies can be also insightful for packaging 
optimization problems \cite{Vetter14,Shaebani17} and has a high potential 
for technological applications. Although interlocking constraints exist 
between polymer chains with irregular structures, our results show that 
the regular gaps between the grains in commercial granular chains can 
significantly affect the strength of entanglements and the rheological 
response of the system. Understanding the interplay of entanglements 
and friction is a key to answer how entangled systems of macroscale 
long objects yield and flow.

M.R.S.\ acknowledges support by the Young Investigator Grant of the Saarland 
University, Grant No.\ 7410110401 and by the Deutsche Forschungsgemeinschaft 
(DFG) through Collaborative Research Center SFB 1027. M.H.\ acknowledges funding 
from the Netherlands Organization for Scientific Research through NWO-VIDI grant 
No.\ 680-47-548/983. 

\begin{widetext}
\noindent {\bf Appendix: Calculation of the mean end-to-end distance of a flexible chain}

\noindent In order to calculate the mean end-to-end distance $\xi$ of a flexible chain 
of length $N$ with bond length $D{+}g$, we first consider the problem in two 
dimensions (see Fig.\,\ref{Fig5}). Assuming that the first grain of the chain is located 
at the origin in an arbitrary Cartesian frame, the angle of the first bond with 
the $x$-axis, $\theta_1$, can be represented as a random angle with an isotropic 
distribution $P(\theta_1){=}\frac{1}{2\pi}$. However, the next successive local 
orientations of the chain are correlated since the next bond angles $\theta_\text{i}$ 
are limited to uniformly vary within the $-\theta_\text{max}\,{\leq}\,\theta_\text{i}
\,{\leq}\,\theta_\text{max}$ range, i.e., the bond-angle distribution is 
$P(\theta_\text{i}){=}\frac{1}{2\theta_\text{max}}$ for $\text{i}{\geq}2$. 
The directional changes of the chain can be modeled as a persistent random walk, 
i.e., a stochastic motion with a preference to move along the arrival direction 
\cite{Tierno16,Nossal74}. We quantify the mean directional persistence of the 
chain with $p\,{=}\,\langle\cos\theta\rangle\,{=}\!\int\!d\theta\,P(\theta) 
\cos(\theta)$, which varies from $p\,{=}\,1$ for a straight rod-like chain 
to lower values ($1\,{<}\,p\,{\leq}\,0$) for flexible chains. 

\begin{figure}[b]
\centering
\includegraphics[width=0.55\textwidth]{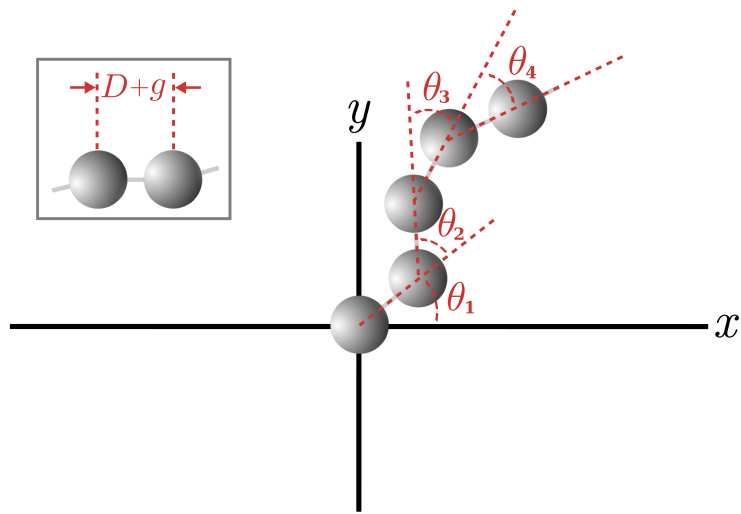}
\caption{Schematic of a few first grains of a granular chain. The initial 
angle with the $x$-axis and the successive bond angles are denoted with 
$\theta_1$ and $\theta_\text{i}$ ($\text{i}{\geq}2$), respectively.}
\label{Fig5}
\end{figure}

By projecting each bond along the $x$-axis, the $x$-coordinate of the $N$-th grain 
center is given as
\begin{equation}
x = (D{+}g) \sum_{\text{i}{=}1}^{N{-}1} \cos\alpha_\text{i},
\label{x-projection}
\end{equation}
where $\alpha_\text{i}$ is the $\text{i}$-th bond angle with the $x$-axis, 
given as $\alpha_\text{i}\,{=}\,\theta_1{+}\cdot\cdot\cdot{+}\,\theta_\text{i}$. 
Using the uniform distributions $P(\theta_1)$ and $P(\theta_\text{i})$ (for 
$\text{i}{\geq}2$), the ensemble-averaged $x$-coordinate of the $N$-th grain, 
$\langle x \rangle$, can be obtained as
\begin{equation}
\begin{aligned}
\langle x \rangle = (D{+}g) \sum_{\text{i}{=}1}^{N{-}1} \!\! 
\int_{-\theta_\text{max}}^{\theta_\text{max}} \!\!\!\!\!\!\!\!\!\!\!\! 
d\theta_\text{i} P(\theta_\text{i}) \cdot\cdot\cdot \!\! 
\int_{-\theta_\text{max}}^{\theta_\text{max}} \!\!\!\!\!\!\!\!\!\!\!\! 
d\theta_2 P(\theta_2) \! \int_{-\pi}^{\pi} \!\!\!\!\!\! 
d\theta_1 P(\theta_1) \cos(\theta_1{+}\cdot\cdot\cdot{+}\theta_\text{i})\!=\! 0,
\label{mean-x}
\end{aligned}
\end{equation}
because the integral over $\theta_1$ vanishes. 

Next, we calculate the second moment of displacement of the $N$-th grain, $\langle 
x^2 \rangle$. Using Eq.\,(\ref{x-projection}), $\langle x^2 \rangle$ reads 
\begin{equation}
\begin{aligned}
\langle x^2 \rangle = (D{+}g)^2\,\Big\langle \!\sum_{\text{i}{=}1}^{N{-}1} 
\sum_{\text{j}{=}1}^{N{-}1} \cos\alpha_\text{i}\,\cos\alpha_\text{j} \Big\rangle 
= (D{+}g)^2\,\Big\langle \!\sum_{\text{i}{=}1}^{N{-}1} \cos^2\!\alpha_\text{i} 
+ 2 \sum_{\text{i}{>}\text{j}} \cos\alpha_\text{i}\,\cos\alpha_\text{j} \Big\rangle.
\label{msd-i}
\end{aligned}
\end{equation}
The ensemble average of the first term leads to 
\begin{equation}
\begin{aligned}
(D{+}g)^2 \!\sum_{\text{i}{=}1}^{N{-}1} \langle \cos^2\!\alpha_\text{i} \rangle 
= (D{+}g)^2 \!\sum_{\text{i}{=}1}^{N{-}1} \!\! 
\int_{-\theta_\text{max}}^{\theta_\text{max}} \!\!\!\!\!\!\!\!\!\!\!\! 
d\theta_\text{i} P(\theta_\text{i}) \cdot\cdot\cdot \!\! 
\int_{-\theta_\text{max}}^{\theta_\text{max}} \!\!\!\!\!\!\!\!\!\!\!\! 
d\theta_2 P(\theta_2) \! \int_{-\pi}^{\pi} \!\!\!\!\!\! 
d\theta_1 P(\theta_1) \cos^2\!(\theta_1{+}\cdot\cdot\cdot{+}\theta_\text{i}) 
= \frac{N{-}1}{2} (D{+}g)^2,
\label{msd-ii}
\end{aligned}
\end{equation}
and the second term of Eq.(\ref{msd-i}) can be evaluated as
\begin{equation}
\begin{aligned}
2(D&{+}g)^2 \sum_{\text{i}{>}\text{j}} \langle \cos\alpha_\text{i}\,
\cos\alpha_\text{j} \rangle \\
&= 2\,(D{+}g)^2 \sum_{\text{i}{=}1}^{N{-}1} \sum_{\text{j}{=}1}^{\text{i}{-}1} 
\int_{-\theta_\text{max}}^{\theta_\text{max}} \!\!\!\!\!\!\!\!\!\!\!\! 
d\theta_\text{i} P(\theta_\text{i}) \cdot\cdot\cdot \!\! 
\int_{-\theta_\text{max}}^{\theta_\text{max}} \!\!\!\!\!\!\!\!\!\!\!\! 
d\theta_2 P(\theta_2) \! \int_{-\pi}^{\pi} \!\!\!\!\!\! 
d\theta_1 P(\theta_1) \cos(\theta_1{+}\cdot\cdot\cdot{+}\theta_\text{i})
\cos(\theta_1{+}\cdot\cdot\cdot{+}\theta_\text{j}) \\
&= (D{+}g)^2 \sum_{\text{i}{=}1}^{N{-}1} \sum_{\text{j}{=}1}^{\text{i}{-}1} 
\int_{-\theta_\text{max}}^{\theta_\text{max}} \!\!\!\!\!\!\!\!\!\!\!\! 
d\theta_\text{i} P(\theta_\text{i}) \cdot\cdot\cdot \!\! 
\int_{-\theta_\text{max}}^{\theta_\text{max}} \!\!\!\!\!\!\!\!\!\!\!\! 
d\theta_{\text{j}{+}1} P(\theta_{\text{j}{+}1}) \cos(\theta_{\text{j}{+}1}
{+}\cdot\cdot\cdot{+}\theta_\text{i}) \\
&= (D{+}g)^2 \sum_{\text{i}{=}1}^{N{-}1} \sum_{\text{j}{=}1}^{\text{i}{-}1} 
\Bigg(\!\int_{-\theta_\text{max}}^{\theta_\text{max}} 
\!\!\!\!\!\!\!\!\!\!\!\! d\theta\,P(\theta)\,\text{Re}[e^{i\theta}]
\Bigg)^{\text{i}{-}\text{j}} \\
&= (D{+}g)^2 \sum_{\text{i}{=}1}^{N{-}1} \sum_{\text{j}{=}1}^{\text{i}{-}1} 
p^{\text{i}{-}\text{j}}=(D{+}g)^2\frac{p}{1{-}p}\big[N{-}1{-}\frac{1{-}
p^{N{-}1}}{1{-}p}\big].
\label{msd-iii}
\end{aligned}
\end{equation}
Similar results as Eqs.\,(\ref{msd-i})-(\ref{msd-iii}) can be obtained for 
$\langle y^2 \rangle$ due to symmetry. Hence, we obtain the following 
expression for the mean end-to-end distance of a flexible chain  
\begin{equation}
\begin{aligned}
\xi = \sqrt{\langle x^2 \rangle + \langle y^2 \rangle} = (D{+}g) \Big[
\frac{1{+}p}{1{-}p}N + \frac{1}{(1{-}p)^2} (2p^N{+}p^2{-}2p{-}1)\Big]^{0.5}
\label{mean-end-to-end}
\end{aligned}
\end{equation}
for $N{>}1$. This equation reduces to $\xi\,{=}\,\sqrt{\frac{1{+}p}{1{-}p}N}(
D{+}g)$ in the large $N$ limit \cite{Doi86}. Although extension of the above 
approach to three dimensions is possible, a more systematic formalism to 
calculate the moments of displacement of a persistent random walk via a 
Fourier-z-transform technique was previously developed in\cite{Shaebani14,
Sadjadi15}. Following this approach, it can be verified that the same 
expression for $\xi$ as for the 2D solution presented in Eq.(\ref{mean-end-to-end}) 
holds in 3D, however, with replacing the previous definition of the 
directional persistence $p$ in 2D with $p\,{=}\!\int\!d\theta\,\sin(\theta)
\,P(\theta) \cos(\theta)$. The approach can be also extended to multi-state 
persistent random walks \cite{Shaebani19}, which is applicable to the 
case of dealing with a non-uniform directional persistence along a 
flexible chain.
\end{widetext}

\bibliography{Refs}

\end{document}